# Five carbon- and nitrogen-bearing species in a hot giant planet's atmosphere

Paolo Giacobbe[1]*, Matteo Brogi[2,1,20], Siddharth Gandhi[2,20], Patricio E. Cubillos[3], Aldo S. Bonomo[1], Alessandro Sozzetti[1], Luca Fossati[3], Gloria Guilluy[1,4], Ilaria Carleo[5], Monica Rainer[6], Avet Harutyunyan[7], Francesco Borsa[8], Lorenzo Pino[9,6] , Valerio Nascimbeni[10], Serena Benatti[11], Katia Biazzo[12], Andrea Bignamini[13], Katy L. Chubb[14], Riccardo Claudi[15], Rosario Cosentino[7], Elvira Covino[16], Mario Damasso[1], Silvano Desidera[15], Aldo F. M. Fiorenzano[7], Adriano Ghedina[7], Antonino F. Lanza[17], Giuseppe Leto[17], Antonio Maggio[11], Luca Malavolta[10], Jesus Maldonado[11], Giuseppina Micela[11], Emilio Molinari[18], Isabella Pagano[17], Marco Pedani[7], Giampaolo Piotto[10], Ennio Poretti[7], Gaetano Scandariato[17], Sergei N. Yurchenko[19], Daniela Fantinel[15], Alberto Galli[6], Marcello Lodi[11], Nicoletta Sanna[6], Andrea Tozzi[6]

**The atmospheres of gaseous giant exoplanets orbiting close to their parent stars (hot Jupiters) have been probed for nearly two decades[1,2]. They allow us to investigate the chemical and physical properties of planetary atmospheres under extreme irradiation conditions[3]. Previous observations of hot Jupiters as they transit in front of their host stars have revealed the frequent presence of water vapour[4] and carbon monoxide[5] in their atmospheres; this has been studied in terms of scaled solar composition[6] under the usual assumption of chemical equilibrium. Both molecules as well as hydrogen cyanide were found in the atmosphere of HD 209458b[5,7,8], a well studied hot Jupiter (with equilibrium temperature around 1,500 kelvin), whereas ammonia was tentatively detected there[9] and subsequently refuted[10]. Here we report observations of HD 209458b that indicate the presence of water ($H_2O$), carbon monoxide (CO), hydrogen cyanide (HCN), methane ($CH_4$), ammonia ($NH_3$) and acetylene ($C_2H_2$), with statistical significance of 5.3 to 9.9 standard deviations per molecule. Atmospheric models in radiative and chemical equilibrium that account for the detected species indicate a carbon-rich chemistry with a carbon-to-oxygen ratio close to or greater than 1, higher than the solar value (0.55). According to existing models relating the atmospheric chemistry to planet formation and migration scenarios[3,11,12], this would suggest that HD

[1] INAF-Osservatorio Astrofisico di Torino, Pino Torinese, Italy. [2] Department of Physics, University of Warwick, Coventry, UK. [3] Space Research Institute, Austrian Academy of Sciences, Graz, Austria. [4] Dipartimento di Fisica, Università di Torino, Torino, Italy. [5] Astronomy Department and Van Vleck Observatory, Wesleyan University, Middletown, USA. [6] INAF-Osservatorio Astrofisico di Arcetri, Arcetri, Italy. [7] INAF-Fundación Galileo Galilei, Breña Baja, Spain. [8] INAF – Osservatorio Astronomico di Brera, Merate, Italy. [9] Anton Pannekoek Institute for Astronomy, University of Amsterdam, Amsterdam, NL. [10] Dipartimento di Fisica e Astronomia "Galileo Galilei", Università di Padova, Padua, Italy. [11] INAF-Osservatorio Astronomico di Palermo, Palermo, Italy. [12] INAF-Osservatorio Astronomico di Roma, Monte Porzio Catone, Italy. [13] INAF-Osservatorio Astronomico di Trieste, Trieste, Italy. [14] Netherlands Institute for Space Research, Utrecht, NL. [15]INAF-Osservatorio Astronomico di Padova, Padua, Italy. [16] INAF - Osservatorio Astronomico di Capodimonte, Napoli, Italy. [17] INAF - Osservatorio Astrofisico di Catania, Catania, Italy. [18] INAF-Osservatorio Astronomico di Cagliari, Selargius, Italy. [19] Department of Physics & Astronomy, University College of London, London, UK [20]Centre for Exoplanets and Habitability, University of Warwick, Coventry, UK. *e-mail: paolo.giacobbe@inaf.it

**209458b formed far from its present location and subsequently migrated inwards[11,13]. Other hot Jupiters may also show a richer chemistry than has been previously found, which would bring into question the frequently made assumption that they have solar-like and oxygen-rich compositions.**

We observed four transits of HD 209458b, the archetype of transiting hot Jupiters, with the near-infrared echelle spectrograph GIANO-B[14], mounted at the 3.6-m Telescopio Nazionale Galileo located in La Palma, Spain. The transits happened on 7 July 2018, 29 August 2018, 27 August 2019 and 3 September 2019. GIANO-B achieves simultaneous coverage in the wavelength range 0.92-2.45 μm, split into fifty orders, at a spectral resolving power of R = 50,000 (ref. [15]). This wide spectral range is key to the detection of multiple molecular species, because their opacity varies widely as a function of wavelength[16].

The raw spectra were optimally extracted using the GOFIO instrument pipeline version 1.6 (ref. [17]), and subsequently calibrated and processed using a custom analysis (see Methods). The main purpose of the latter is to refine the wavelength calibration of the spectra, changing both from night to night and during the same night, and to remove the unwanted spectral absorption lines formed in the Earth's atmosphere (telluric absorption) and in the stellar photosphere. One important aspect of these observations is that, during a transit event, the planet moves along the orbit, thus producing a change in its radial velocity between −16 km s$^{-1}$ at the transit ingress and +16 km s$^{-1}$ at the transit egress. In contrast, the telluric and stellar spectra are nearly stationary and can be effectively filtered out by our pipeline.

Each order of each processed spectrum was first cross-correlated with transmission models computed over the spectral range of GIANO-B by assuming an isothermal atmosphere and constant volume mixing ratios (VMRs) for the seven major species driving the chemistry of hot Jupiters[6] ($H_2O$, HCN, $CH_4$, $NH_3$, $C_2H_2$, CO and $CO_2$). Temperatures and VMRs were chosen within the ranges $1{,}000 < T < 1{,}500$ K and $10^{-5} < \text{VMR} < 10^{-2}$, and the models convolved with the instrument profile (see Methods). Although these values do not match any specific chemical scenarios, they were used to maximize the detection significance without any assumptions about the chemical composition of the atmosphere. For each species tested, we applied an automated procedure to select the orders of the spectrograph to use for cross-correlation (see Methods). These are typically the orders that contain the strongest spectral lines of the planet spectrum and are less contaminated by absorption in the Earth's atmosphere. The cross-correlation functions (CCFs) from the selected orders were then co-added in time with equal weighting as a function of planet radial velocity and maximum radial velocity semi-amplitude $K_P$ (see Methods).

From cross-correlation with the isothermal models, we detect the signal of six of the seven species tested, namely $H_2O$ (9.6σ), HCN (9.9σ), $C_2H_2$ (6.1σ), CO (5.5σ), $NH_3$ (5.3σ), and $CH_4$ (5.6σ). We do not detect $CO_2$ with confidence level higher than 3σ. The significance maps are shown in Fig. 1. For all species, we calculated the detection significance by performing a Welch t-test[18] on two samples of cross-correlation values: the former far from the planet radial velocity (>25 km s$^{-1}$), the latter near to it (<3 km s$^{-1}$). We reject the null hypothesis that the two samples have the same mean, and transform the corresponding P value into significance through a two-tailed test, as correlation values can be positive and negative (see Methods and Extended Data Fig. 1).

Given the strength of the signal from $H_2O$ and HCN, we can detect them individually in each of the four transits. The other four species have weaker signatures and are not always detected above the threshold of 3σ per transit. However, they are firmly detected when multiple transits are co-added. Furthermore, for all species, the detection significance always increases with increasing number of co-added transits. This suggests that all species are present in all observing nights, although the data quality does not allow us to confirm or exclude the presence of night-to-night variability. Overall, these results show the advantage of a multi-night approach, which allows us to detect the atmosphere of HD 209458b consistently, supporting the genuine nature of the measured signal. We also conducted a series of tests to assess whether cross-correlation techniques can reliably extract information from spectra with so many mixed species (see Methods for the details).

To interpret the physical and chemical conditions compatible with the simultaneous detection of six species in the atmosphere of HD 209458b, we computed two sets of non-isothermal atmospheric models. For the first set, we used input temperature-pressure-abundance profiles calculated under the assumptions of a cloud-free atmosphere in chemical and radiative equilibrium (see Methods). In these models, we explored overall atmospheric elemental metallicity ranging between 0.001× solar and 100× solar, in steps of one decade (six values), and carbon-to-oxygen ratio (C/O) values of 0.1, 0.55, 0.90, 0.95, 1.05, 1.5 and 2.0. For each chemical scenario, the temperature-pressure-abundance profiles were allowed to adjust self-consistently (see Methods for the details and Extended Data Figs. 2, 3). The second set of models accounts for the presence of clouds/aerosol by adding a grey cloud deck with a top-deck pressure of $10^{-5.5}$ bar and a cloud fraction of 0.4 (refs. [10,19,20]). Our cloud prescription does not take possible elemental sequestration via condensation into account. Therefore, it does not alter either the radiative equilibrium of the planetary atmosphere or the atmospheric C/O ratio.

Since the abundance, and thus detectability, of nitrogen (N)-bearing and some carbon (C)-bearing species could be increased by processes out of thermochemical equilibrium, we also tested a disequilibrium scenario for solar composition. For this purpose, we considered specific temperature-pressure-abundance profiles for HD 209458b at the terminator average, accounting for photochemistry and transport disequilibrium processes[21]. These processes could yield higher abundances for HCN, $NH_3$, $CH_4$ and possibly $C_2H_2$ (ref. [22]), than predicted for a solar-composition in thermochemical equilibrium.

We cross-correlated the grids of equilibrium and disequilibrium models with the GIANO-B spectra and converted the cross-correlation values into likelihood values[23] (see Methods). We then used a likelihood-ratio test to compare the different models, by taking for each of them the maximum likelihood at the expected planet radial-velocity semi-amplitude $K_P$.

Our results statistically favour the presence of aerosols in the atmosphere of HD 209458b (Extended Data Fig. 4), which dampen the amplitude of the molecular lines but do not evidently hamper their detection[24,25]. This supports the results of other observations of this planet (see Methods). By considering the models with all the species mixed, the planet metallicity is not well constrained as values from 0.001× solar to 10× solar are consistent within 2σ, though highly sub-solar metallicities of 0.001-0.01× solar would be marginally favoured (Fig. 2). For metallicities higher than 0.1× solar, C/O ratios greater than the solar value, that is, C/O ≳ 0.9, are statistically favoured at more than 4σ, while a wider range of C/O down to 0.5 would be in principle possible at the lowest metallicity considered (0.001 × solar). However, this metallicity is not supported by the confidence intervals of the species HCN and $C_2H_2$ showing a strong preference for metallicities from 0.1× to 10× solar and C/O ≳ 1 (Fig. 3). This ambiguity can be explained by the fact that the models with the mixed species are mainly affected by the opacity of water vapour, which dominates over the other species (for example, HCN, $C_2H_2$). Indeed, at highly sub-solar metallicities, the posterior distribution of the mixed models mainly reflects that of water (see Fig. 2 and Fig. 3). For this reason, we argue that atmospheric C/O ≳ 1 ratios and 0.1×-10× solar metallicities, as indicated by HCN, $C_2H_2$, and, to a lesser extent, CH4, are the most likely for HD 209458b ($NH_3$ and CO abundances are almost insensitive to the C/O value).

We note that a unity C/O ratio is a tipping point in equilibrium chemistry because it marks the transition at which the dominant molecules in the atmosphere shift from oxygen-rich to carbon-rich species. Over the parameter space explored, carbon monoxide is normally the dominant carbon-bearing species. For C/O < 1, the formation of CO is limited by the amount of available carbon (that is, there is less carbon than oxygen available, and thus, CO consumes most of the available carbon). Vice versa, for C/O > 1 the formation of CO is limited by the amount of available oxygen. The excess available carbon triggers the formation of other carbon-bearing species, such as HCN, $CH_4$ or $C_2H_2$, thus raising their abundances by orders of magnitude compared to the scenario with C/O < 1.

The tested atmospheric models in thermochemical disequilibrium are strongly disfavoured with respect to the equilibrium models at more than 16σ and 30σ for the cloudless and cloudy models, respectively. This is mainly due to the fact that the disequilibrium scenarios for solar composition predict a much higher abundance of water vapour than our observations can account for. Nevertheless, we cannot exclude that disequilibrium processes may take place to some extent. A more sophisticated treatment of disequilibrium chemistry for different compositions, possibly using the most recent three-dimensional atmospheric models[26], will be required to explore more in depth out-of-equilibrium scenarios.

Under the assumption of the validity of thermochemical equilibrium, our estimate of the atmospheric C/O ratio may yield constraints on the formation and migration processes of the hot Jupiter HD 209458b, relying on theoretical frameworks developed to date[3]. Specifically, a C/O ratio close to 1, compatible with our data, would indicate that HD 209458b formed beyond the $H_2O$ condensation front (snowline) at about 2-3 au, more likely between the snowlines of $CO_2$ at 5-8 au and CO at about 30-40 au, and then migrated inward to its current orbital separation at 0.047 au with no substantial accretion of oxygen-rich solids or gas[11,12,27] (see Methods). However, our estimate of C/O ratio does not take into account possible rainout of oxygen-rich refractory species[28], which might increase C/O ratios slightly lower than 1 (around 0.8-0.9) to the measured atmospheric C/O ≳ 1, but whose impact cannot be properly assessed in this work.

Twenty years after the observation of the first transiting planet HD 209458b, the detection of six molecular species in its hot atmosphere at high resolution reveals a carbon-rich and complex chemistry. Future observations at low resolution in relatively wide near-infrared and mid-infrared spectral bands with the James Webb[29] and ARIEL[30] space telescopes, their combination with current and future high-resolution spectra, and new developments in atmospheric modelling and retrieval, are required to further characterize the atmosphere of HD 209458b and to investigate the atmospheres of other hot Jupiters with similar C/O ≳ 1. This will allow us to study the chemistry of carbon-rich atmospheres in both thermochemical equilibrium and disequilibrium more thoroughly than we could do here. The same approach in the data analysis used in this work will enable us to study the atmospheres of smaller and cooler exoplanets, even in the habitable zone, as soon as the required instrumentation at high resolution becomes available.

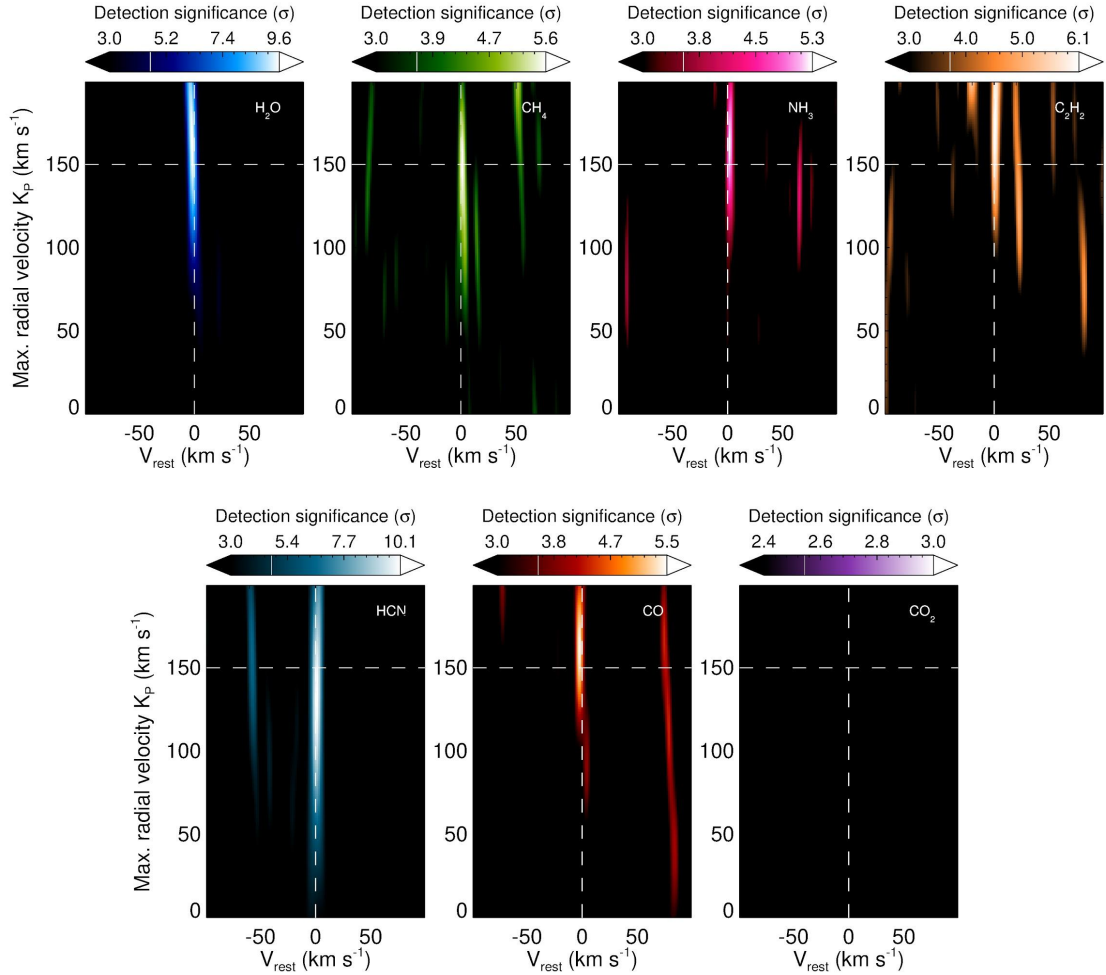

**Fig. 1 | Detection significance for $H_2O$, $CH_4$, $NH_3$, $C_2H_2$, HCN, CO and $CO_2$.** Each panel shows the significance of the cross-correlation of the GIANO-B spectra with isothermal atmospheric models, as a function of the planet's maximum radial velocity ($K_P$) and the planet's rest-frame velocity ($V_{rest}$). White dashed lines denote the known velocity of HD 209458b, that is ($K_P$, $V_{rest}$) = (145, 0) km s$^{-1}$. The significance was computed with a Welch t-test on two samples of cross-correlation values, that is, far from and near to the planet radial velocity respectively (see text).

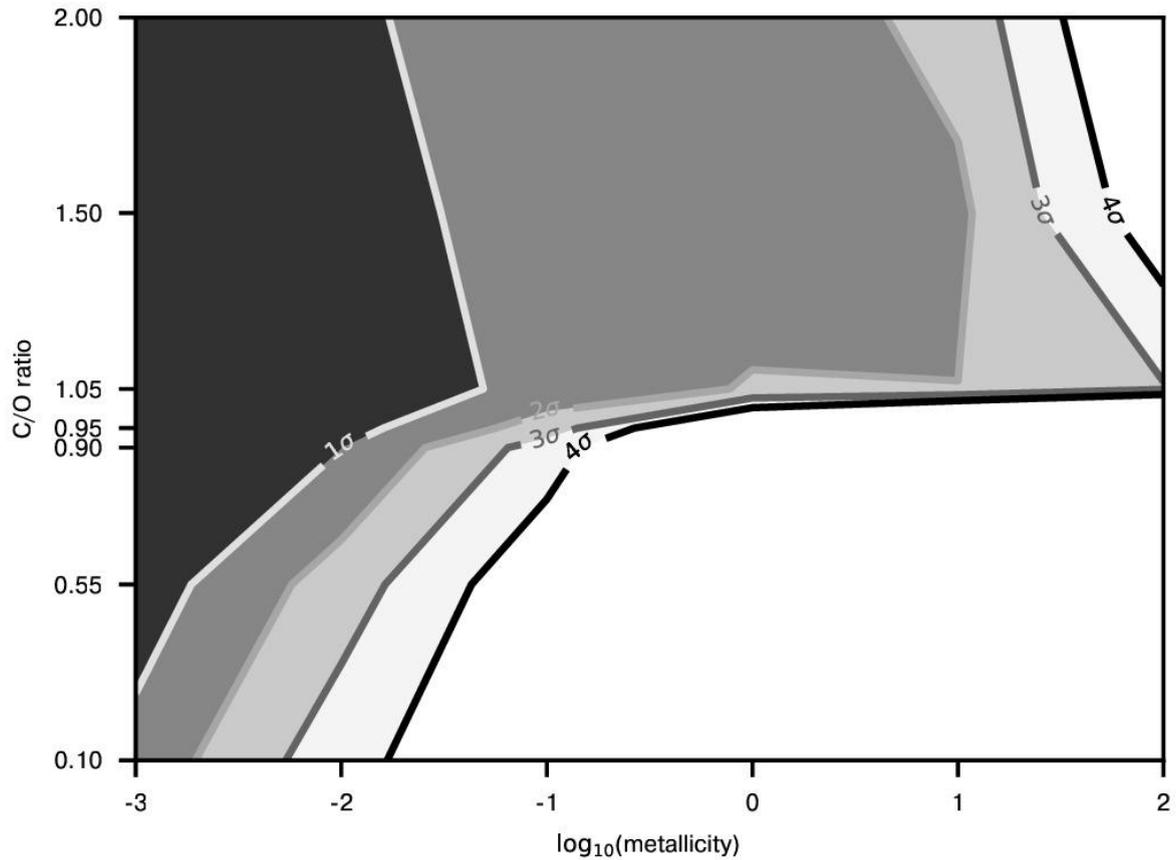

**Fig. 2 | Constraints on the chemical composition of the atmosphere of HD 209458b.** Confidence intervals of the C/O ratio (vertical axis) and metallicity (horizontal axis) are shown between 1σ and 4σ, for an atmosphere in thermochemical equilibrium with clouds, by considering the detected molecular species all together. They are derived from likelihood-ratio tests after converting the cross-correlation coefficients to likelihood values (see Methods).
Similar results are found for the clear models with no clouds (see Extended Data Fig. 4).

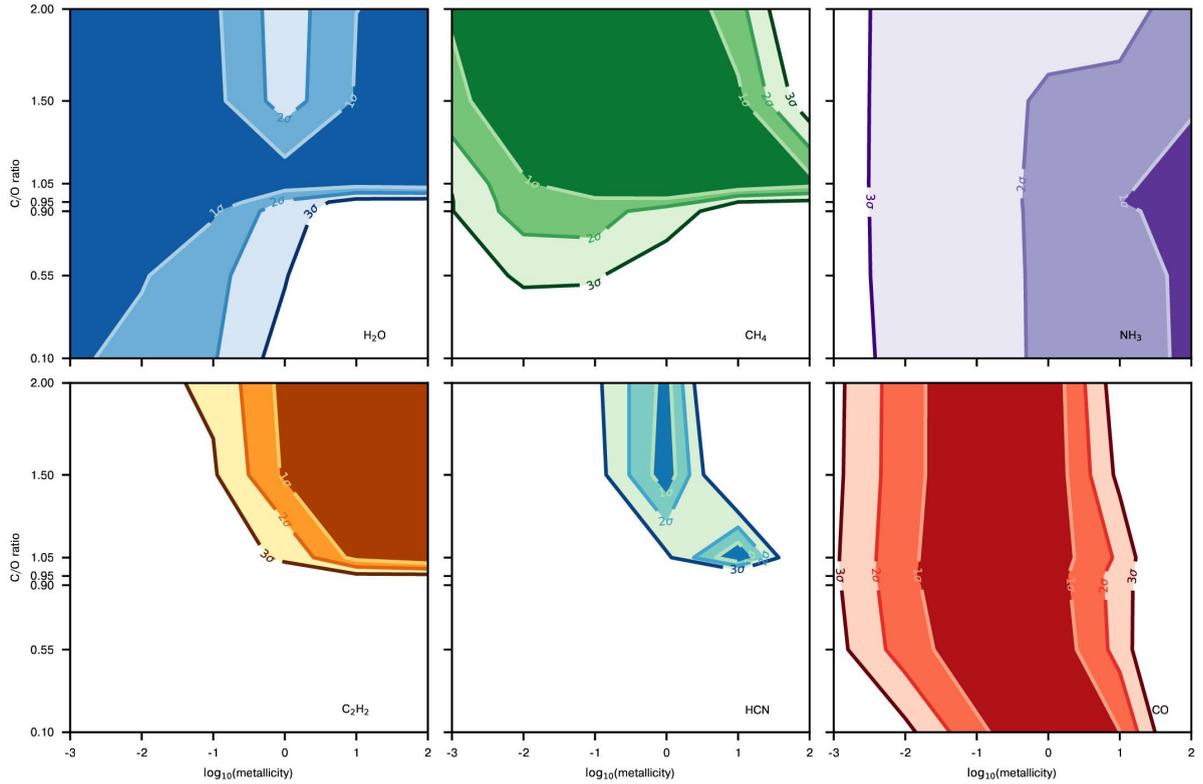

**Fig. 3 | Single-molecule constraints on the chemical composition of the atmosphere of HD 209458b.** Confidence intervals of the C/O ratio (vertical axis) and metallicity (horizontal axis) are shown between 1σ and 3σ, for an atmosphere in thermochemical equilibrium with clouds, by considering the detected molecular species ($H_2O$, $CH_4$, $NH_3$, $C_2H_2$, HCN and CO) one at a time. They are derived from likelihood-ratio tests after converting the cross-correlation coefficients to likelihood values (see Methods). Similar results are found for the clear models with no clouds (not shown here).

**METHODS**

**Observations and data reduction**

We observed five transits of HD 209458b as part of the GAPS project[31] during a Long-Term programme (Principal Investigator: G.M.) with the near-infrared high-resolution GIANO-B spectrograph[15], mounted at the Telescopio Nazionale Galileo. The observations were carried out in GIARPS (GIANO-B + HARPS-N; ref. [14]) configuration mode and were performed with the nodding acquisition mode ABAB, where target and sky spectra were taken in pairs while alternating between two nodding positions along the slit (A and B) separated by 5″. This enables an optimal subtraction of the detector noise and background. Further details about the observational strategy, substantially unchanged from previous works, can be found in ref. [32] and ref. [33]. We collected a total of 276 spectra with an exposure time of 200 s per spectrum. All the observations were scheduled in order to obtain spectra pre-, after- and in-transit with airmass between 1 and 2. We measure a mean signal-to-noise ratio (SNR) between ~80 and ~120 per spectrum per pixel averaged across the entire dataset and the entire spectral range. The observations log, night by night, is provided in Extended Data Table 1. In the next steps of the analysis, we do not consider the night of 5 September 2018, because about half of the transit observations were lost owing to the presence of clouds.

GIANO-B spectra cover the *Y*, *J*, *H* and *K* spectral bands (0.95-2.45 μm) in 50 orders at a mean spectral resolving power of $R \approx 50{,}000$. The raw spectra were dark-subtracted, flat-corrected, and extracted using the GOFIO pipeline version 1.6 (refs. [17,34]) without applying the blaze function correction. Although GOFIO also performs a preliminary wavelength calibration using U-Ne lamp spectra as a template, the mechanical instability of the instrument causes the wavelength solution to change during the observations. Since the U-Ne lamp spectrum is only acquired at the end of the observations in order to avoid persistence, the solution determined by GOFIO is not sufficiently accurate, and in fact, we expect the wavelength solution of the spectra to shift and jitter between consecutive exposures[32]. We correct for this jitter by aligning all the observed spectra to the telluric reference frame via cross-correlation with a time-averaged observed spectrum of the target used as a template. The details of the procedure are described in previous work[33]. We achieve a residual scatter in the measured peak position of the CCF well below 0.1 km s$^{-1}$, that is, approximately 1/30th of a pixel. The telluric spectrum also provides an excellent wavelength calibration source. We apply the same procedure as in past work[32,33] to refine the standard GOFIO wavelength calibration. It consists in matching a set of telluric lines in the time-averaged observed spectrum with a high-resolution model of the Earth transmission spectrum generated via the ESO Sky Model Calculator[35], and solve for the (pixel, wavelength) relation with a fourth-order polynomial fit. As in previous work, this calibration approach is not possible for the orders that show few (or no) telluric lines or for the orders where Earth's atmosphere is particularly opaque (that is, with heavily saturated spectral lines). In the former case, we use the standard wavelength calibration based on the U-Ne lamp, whereas in the latter case we discard the saturated orders. After this selection, we exclude orders 8, 9, 10, 23, 24, 30, 45, 46, 47, 48 and 49 from our analysis (Extended Data Fig. 5, grey bands). We point out that, in the GIANO-B spectra, order 0 is the reddest and order 49 is the bluest. The precision of our calibration is estimated by computing the standard deviation of the residuals after fitting the (pixel, wavelength) relation for each of the telluric lines. We achieve a residual scatter per line well below 1 km s$^{-1}$, that is, approximately one-third of a pixel.

**Telluric removal via principal component analysis (PCA)**

At this stage of the analysis, the planet's transmission spectrum is completely overshadowed by the telluric and stellar spectra. After re-aligning the spectral sequence as above, we can use the fact that the transmission spectrum of the Earth and the stellar spectrum are stationary (or quasi-stationary considering that the stellar barycentric velocity changes by about 18 m s$^{-1}$ during transit) signals in wavelength, so their spectral lines will always fall on the same pixel. In contrast, the planetary absorption spectrum is Doppler-shifted by tens of kilometres per second due to the variation of the orbital radial velocity of the planet. At the resolution of GIANO-B, this shift corresponds to around 10 pixels on the detector during transit. As in previous high-resolution spectroscopy, we exploit this peculiar Doppler signature to remove the telluric and stellar spectra while preserving the planet signal.

In this work we design a PCA with which to analyse GIANO-B spectra, but the algorithm described below can be applied to any high-resolution observations. PCA has been successfully applied in the past on VLT/CRIRES[36,37] and Keck/NIRSPEC[38,39] data to remove the telluric spectrum. The idea behind the method is to consider any correlated inhomogeneities in the wavelength or temporal domain (for example, the stellar spectrum, the blaze function of the instrument, the telluric absorption as a function of the airmass, and any correlated noise) as a systematic trend shared by each spectrum during the transit observations. Thus the aim of PCA is to find a basis of representative trends that describe, via linear combination, all the varying signals in our data. PCA works on the covariance matrix between datasets and computes the principal components (or eigenvectors) of an $M \times N$ (columns × rows) matrix, where $M$ is the number of variables and $N$ is the number of observations or samples. The calculated set of eigenvectors, associated with their eigenvalues, fully describe the initial covariance matrix. From an algebraic point of view, PCA describes the initial matrix in a new orthogonal reference system and therefore computes as many eigenvectors as the number of samples considered.

With our spectroscopic dataset, two configurations are possible to define samples and variables. The first of such configurations is the matrix that describes the time domain, where the single spectra are the rows and the wavelength channels are the columns. The alternative configuration is the wavelength domain, where we transpose the time-domain matrix to have the spectra as columns and wavelength channels as rows. The two domains are sensitive to different effects, for example, the telluric plus stellar spectrum in the wavelength domain or the airmass variation in the time domain, and indeed the PCA technique has previously been successfully applied in both domains. Regardless of the choice, since the number of observed spectra is much smaller than the number of wavelength channels, we obtain rectangular matrices of very unequal dimensions. The dimension of the covariance matrix and the number of eigenvectors are equal to the number of rows of the input matrix, so this number will change whether we adopt the time-domain or wavelength-domain representation of the data. However, we note that the number of free parameters (that is, the number of components) that we can use to filter our spectra is always determined by the smaller dimension (that is, the number of spectra). For the rest of the discussion, we explain each step of the filtering technique considering only the wavelength-domain representation and a single spectral order. The same procedures apply to all the other orders without variations.

Before computing the PCA we perform the following steps on the data. Each spectrum (each row) is normalized by its median value. This operation is done to correct baseline flux differences between the spectra that are due, for example, to variable transparency, imperfect telescope pointing, or instability of the stellar point spread function, all of which change the overall amount of flux reaching the detector. Subsequently, each spectral channel (each column) has its mean subtracted. Lastly, each spectrum (each row) is divided by its standard deviation. This procedure is used to 'standardize' the data, that is, to reduce them to a variable with zero mean and unit standard deviation. The result of the previous three steps is an $M$-column, $N$-row matrix $S$, where $M$ is the number of spectral channels (2,048 for a single GIANO-B order) and $N$ is the number of spectra. On this matrix, we apply PCA by using the IDL PCOMP function. We input the covariances of the original data as computed by the IDL CORRELATE function. The output of PCOMP is also an $M$-column, $N$-row matrix where each row contains an eigenvector $E$ of length $M$. Following the conventions of previous work[40], we define a filter function matrix $F$ as:

$$F = \begin{pmatrix} E_{1,1} & \cdots & E_{1,M} \\ \cdots & E_{n,j} & \cdots \\ E_{N_{opt},1} & \cdots & E_{N_{opt},M} \\ 1 & \cdots & 1 \end{pmatrix}$$

where $E_{n,j}$ is the $j$th element of the $n$th eigenvector with the index $n$ running over the optimal number of components $N_{opt}$ and the index $j$ running over the number of spectral channels $M$. Formally, each observed spectrum $S_i$ can be written as a linear combination

$$S_i = F\, c_i + \text{noise}$$

where $S_i$ is the $i$th spectrum with the index $i$ running over the total number of spectra, $c_i$ is the $i$th linear coefficient and $S_i = [S_{i,0} \ldots S_{i,M}]^T$ is the list of the $M$ spectral channels of the $i$th spectrum.

Assuming that the uncertainties on $S_i$ are Gaussian and constant, the maximum likelihood solution for $c_i$ is $\hat{c}_i = (F^T F)^{-1} F^T S_i$. In our code, we use the IDL function SVSOL with the back-substitution technique to solve the linear equations $F\, c_i = S_i$.

To select the appropriate number ($N_{opt}$) of eigenvectors $E$ to be used to build the filter function $F$, we perform an iterative procedure that progressively increases $N_{opt}$ starting from one. For each number of eigenvectors, we perform the linear regression as described above and we use the standard deviation of the $M \times N$ filtered matrix as a quality estimator of the filtering. Considering that the eigenvectors are sorted based on their contribution to the initial variance, the first derivative of the function that describes the standard deviation versus the number of considered eigenvectors tends to zero. We select the optimal number of eigenvectors $N_{opt}$ as the number of eigenvectors for which the first derivative between the last and penultimate component decreases by less than $\sigma_{white} N^{-1/2}$, where $\sigma_{white}$ is the standard deviation of the full matrix assuming pure white noise. Typically, in our data, the limit on the first derivative is about 0.001. This procedure results in an optimal number of components $N_{opt}$ between 2 and 7 depending on the quality of the night and the spectral order.

As a last step of the analysis, we divide each spectral channel by its variance and we multiply the final matrix by the median of the variances, in order to conserve the flux. This ensures that each spectral channel is equally weighted when the transmission spectrum is extracted via CCF (see the section below). The steps described in this section are graphically shown in Extended Data Fig. 6.

**Generation of theoretical transmission spectra**

Theoretical transmission spectra of HD 209458b are computed using the GENESIS model[41] adapted for transmission spectroscopy[42]. Models are calculated between 100 bar and $10^{-8}$ bar in pressure, and between 0.9 μm and 2.6 μm in wavelength, at a constant wavenumber spacing of 0.01 cm$^{-1}$. This choice corresponds to resolving powers between 380,000 and 1,110,000. GENESIS takes in input any temperature-pressure and abundance-pressure (VMR-pressure) profiles. In this study we explore either isothermal models (constant temperature-pressure and VMR-pressure) or models with temperature-pressure and VMR-pressure profiles resulting from both the equilibrium calculations described in the next section and disequilibrium chemistry in ref. [21].

We use the most up-to-date molecular line lists to compute the opacity of the seven species investigated. These prescriptions are currently the most suitable for high-resolution spectroscopy[16]. We use the ExoMol database[43–46] for $H_2O$, $NH_3$, HCN and $C_2H_2$, the HITEMP database[47–49] for $CH_4$ and CO, and the Ames database[50] for $CO_2$. Each spectral line is broadened by pressure and temperature, resulting in a Voigt line profile as a function of frequency[42]. For each species, we use the latest $H_2$ and He pressure broadening coefficients to accurately determine the cross-section in the $H_2$/He rich environment[16]. We additionally include collision-induced absorption from $H_2H_2$ and $H_2$He interactions[51].

Retrieval of optical and infrared low-resolution spectroscopy, mostly informed by the slope of the optical transmission spectrum, the relatively weak alkali lines (sodium and potassium), and the amplitude of the water band around 1.4 μm, are all indicative of some aerosol coverage[4,10,52] in the atmosphere of HD 209458b. Further evidence comes from the detection at near-ultraviolet wavelengths of iron and the non-detection of magnesium in the planetary upper atmosphere[53], which suggests the presence of magnesium-bearing aerosols[54]. To account for aerosols, we include a grey opacity due to a cloud deck by calculating a transmission spectrum with a cloud top pressure of $10^{-5.5}$ bar and weighting this with a clear atmosphere with a cloud fraction of 0.4. These prescriptions for clouds are consistent with constraints from previous analyses of low-resolution data[10,19,20].

We confirm our detections with alternative high-resolution line lists for a number of species, as detailed in Extended Data Table 3. Such tests are crucial for robust detections as recent work has shown that high-resolution detections can be affected by the choice of line list[23,55]. We detect $H_2O$ with both the HITEMP and POKAZATEL line lists at similar significance, but our $CH_4$ and $NH_3$ detections are weaker with the HITRAN line list. Our $C_2H_2$ detection is also confirmed with the ASD-1000 and HITRAN line lists, but the latter results in the weakest detection as it is a room-temperature line list and therefore less complete at the much higher temperatures considered here. We have also tested $CO_2$ with the HITEMP as well as the Ames line lists but we are unable to detect the species with either.

**Temperature-pressure-abundance profiles in thermochemical radiative equilibrium**

To interpret the simultaneous detection of six species on HD 209458b, we drive the radiative transfer calculations described above with temperature-pressure and temperature-VMR profiles modelled in a cloud-free approximation, and with a range of elemental compositions. These are parameterized by metallicity (from 0.1× to 100× solar) and C/O ratio (0.1 to 2.0, where the solar value is C/O ≈ 0.55). These calculations assume local thermodynamic, radiative, hydrostatic and thermochemical equilibrium. They are obtained with a modified version of the Pyrat Bay atmospheric modelling framework[26,56], which implements a one-dimensional two-stream radiative transfer scheme[57]. Abundances are computed in thermochemical equilibrium via the open-source TEA code[58] for any given temperature and elemental composition. The choice of opacities is consistent with the GENESIS code $H_2O$, $NH_3$, HCN, $C_2H_2$, CO but differs for $CO_2$ (HITEMP[47]) and $CH_4$ (ExoMol[59]). Additionally, the radiative transfer accounts for Na and K resonance-line opacity[60], Rayleigh opacity from $H_2$, H and He[61], and collision-induced absorption from $H_2$-$H_2$ and $H_2$-He interactions[62–66]. The other difference from GENESIS is that before the radiative-transfer calculations we process the opacities from ExoMol with the Repack code[67]. This code extracts the dominant line transitions in the wavelength and temperature range of interest, thus reducing the number of transitions from billions to millions by discarding lines that are too weak to contribute to the energy balance of the planet. To attain radiative equilibrium we follow an iterative scheme[57], where we compute the radiative transfer between 0.3 μm and 33 μm, updating the temperature profile towards local energy conservation at each layer after each iteration. Additionally, we update the composition to the thermochemical-equilibrium value according to the current temperature profile every five iterations. After about 100 iterations, the atmosphere converges to a stable temperature profile. The resulting temperature-pressure profiles are shown in Extended Data Fig. 2. The resulting relative abundances vary substantially across the grid (see Extended Data Fig. 3), with a particularly abrupt transition at C/O ≈ 1.0, expected as the dominant elemental species changes from carbon to oxygen. This behaviour is well in agreement with other previous results from the literature[21].

**Extraction of the signal by cross-correlation**

After the removal of stellar and telluric lines via PCA, the residual spectra (lower panel of Extended Data Fig. 6) contain only the exoplanet signal, albeit deeply buried in the noise (that is, individual planet lines have SNR ≪ 1). However, there are thousands of strong molecular lines in the GIANO-B spectral range, and therefore we can combine their signals to attempt a detection of the planet signature. This is done by cross-correlating the residual GIANO-B spectra with models of the planet's transmission spectrum computed as explained previously. In the limit of uncorrelated noise, the final precision on the CCF mainly depends on the number of matched lines. The cross correlation technique is applied to GIANO-B data similarly to past work[32,33]. As a first step, the CCF is computed on a fixed grid of radial velocity lags between −225 km s$^{-1}$ and +225 km s$^{-1}$, in steps of 1.5 km s$^{-1}$ and then re-sampled to 2.7 km s$^{-1}$ (167 values) in order to match the pixel scale of GIANO-B. This operation is done in order to avoid the presence of correlated signals in the CCF due to oversampling. For each radial velocity lag, shifting and re-sampling is obtained via spline interpolation.

CCFs are calculated for every molecule, every spectral order ($N_{ord}$), every exposure ($N_{spectra}$), and every night. The output is therefore a matrix of cross-correlation values with dimensions $167 \times N_{ord} \times N_{spectra}$ for each observing night and each molecule. The CCFs from all the orders selected (see below) are then co-added, resulting in a matrix with dimension $167 \times N_{spectra}$. When co-adding each CCF is equally weighted, even though the signal per order will arguably depend on many factors, such as the transparency of the Earth's atmosphere, the efficiency of the instrumental setup, the density and depth of absorption lines in the transmission spectrum, and the position of exoplanetary lines compared to strong telluric/stellar lines. This is why an optimal selection of the spectral orders, which might differ for each molecule and night by night, has a key role. We will describe our strategy for an optimal order selection in the next section.

Even at this stage, the planet signal is not expected to be detectable in the CCF of the single spectrum. Therefore, we co-add all the spectra obtained during one or multiple transits. This requires shifting the CCFs to the rest-frame of HD 209458b. We compute the planet's radial velocity $V_P$ in the telluric reference system. As the orbit of HD 209458b is circular[68], this is given by:

$$V_P(t) = V_{sys} + V_{bary} + K_P \sin[2\pi\varphi(t)],$$

where $V_{sys}$ is the systemic velocity of the star-planet system with respect to the barycentre of the solar system, $V_{bary}$ is the velocity of the observer induced by rotation of the Earth and by the motion of the Earth around the Sun (that is, the barycentreEarth radial velocity), $\varphi(t)$ is the orbital phase of the planet at time t, and $K_P$ is the planet's orbital radial velocity semi-amplitude. $\varphi(t)$ is obtained as the fractional part of $(t - T_c)/P_{orb}$, where $P_{orb}$ is the orbital period and $T_c$ is the time of mid-transit. Extended Data Table 2 lists the reference values for these calculations.

Although $K_P$ is well constrained by radial velocities and transit measurement to $(145 \pm 1.5)$ km s$^{-1}$, we still explore a full range of values between 0 km s$^{-1}$ and 200 km s$^{-1}$ in steps of 3 km s$^{-1}$. Exploring a sufficiently large parameter space offers a strong diagnostic on all sources of noise and allows us to verify that no other spurious signal produces a significant detection near the planet's rest frame position. For each value of $K_P$, we re-align the CCFs in the planet rest frame (Extended Data Fig. 1b) via linear interpolation and we co-add them in phase. This step maximizes the planetary signal as a function of the rest-frame velocity $V_{rest}$ and the planetary semi-amplitude $K_P$.

**Optimal selection of spectral orders for each molecule**

This procedure consists in selecting the orders where the density and depth of absorption lines are sufficient to substantially contribute to the planet signal, and where the telluric residual signal does not interfere with the planetary signal. The interplay between telluric and the planetary lines is not straightforward to estimate. It changes for each molecule and varies night by night, both in strength (for example, the telluric spectrum depends on the humidity and airmass) and in wavelength position (since the barycentric velocity varies night by night). Furthermore, the efficiency of GIANO-B also varies as a function of wavelength.

To account for all the effects listed above, our approach relies on injecting model planetary spectra into the observations just before the PCA procedure. We then recover these artificial signals and measure their significance order by order, by using the same procedure as for the molecule detection. The injected models are computed as those used for cross-correlation, except that they are amplified to be easily detectable in the best spectral orders, but not at significances larger than $6\sigma$-$10\sigma$. This ensures that the overall level of the injected signal is still comparable to the observed signal. An order is selected when the most significant signal is recovered within 6 km s$^{-1}$ from the systemic velocity of the injection and within 30 km s$^{-1}$ of the injected $K_P$ with a significance greater than $3\sigma$. To prevent any interference with the real signal present in the data, we test multiple injections at slightly different velocity positions. We note, however, that when the injected model is amplified by a factor of about 2 or greater, the influence of the real signal becomes negligible and does not alter the order selection. We highlight that our procedure for order selection is fundamentally different from a weighting procedure. We only aim at selecting the orders where molecular absorption is likely to occur, and once we select a set of orders for one species, we keep the same selection for all the models tested, with equal order weighting. Furthermore, we note that we do not attempt to optimize the number of PCA components via maximization of an injected signal.

Although the procedure above can fully account for the effect of telluric lines, there is also the possibility that each molecular species in the exoplanet spectrum is influenced by the complex spectrum of all the other species. To exclude this possibility, we repeated the order selection by injecting a mixed model, that is, a model including the six detected species, but cross correlating with single-species models. We obtained the same order selection for the dominant species ($H_2O$ and HCN), and minor differences for the other species. In none of the cases, however, was the detection significance strongly altered by the chosen procedure. Considering that the model with all species strongly depends on the chosen relative VMRs and that this dependence can lead to biases, we adopt the order selection obtained by using a single-species model, which is shown in Extended Data Fig. 5.

The order selection explained above is not applied to cross-correlation with the mixed models. In this case, we use all the orders of the spectrographs that were successfully calibrated. As indicated in Extended Data Fig. 5, this corresponds to 90% of the available spectral range.

**Determining the significance of the detection**

We determine the significance level of the signal via a Welch $t$-test on two distributions of cross-correlation values, one within $\pm 3$ km s$^{-1}$ of the planet's rest-frame velocity and one more than 25 km s$^{-1}$ away from the planet's rest-frame velocity. An example of the two distributions is shown in Extended Data Fig. 1c. The null hypothesis is that the two samples have the same mean, and the test rejects this hypothesis at a certain significance level, which we adopt as the significance of the detection. We use the IDL function TM_TEST to compute the $t$ value and the relative $P$ value, which is the probability of obtaining a $t$ value greater than the measured value. The corresponding $P$ value is then converted into significance by calculating the inverse survival function of a normal distribution via the Python routine scipy.stats.isf(). The argument of the function is the $P$ value divided by two, which means we perform a two-tailed test because cross-correlation values can deviate both positively and negatively. Extended Data Fig. 1a shows the two-dimensional significance map, where each value of the two-dimensional grid in $K_P$ and $V_{rest}$ is computed as described above. From the map we define $1\sigma$ errors on $K_P$ and $V_{rest}$ as the values for which the significance drops by 1

(Extended Data Fig. 1d, e). Our significance calculations are based on the strong hypothesis of uncorrelated noise, which has been shown to be a valid approximation in past works[32,33] as the distribution of cross-correlation values is indeed Gaussian down to the testable limits given the sample size (typically 3 to 4 standard deviations away from the mean of the distribution).

**Additional tests on the reliability of cross-correlation techniques**

Given that this is the first time that more than two molecular species are measured simultaneously in an exoplanet spectrum, we conducted a series of additional tests to assess whether cross-correlation techniques can reliably extract information from spectra with many mixed species. The main goal of these tests is to prove that it is possible to detect up to seven different species with our data-analysis pipeline with negligible false positives.

As a first test, we show that individual species can be detected even when mixed with the dense forest of lines from other species. We construct synthetic datasets mimicking as closely as possible the real data, in particular their variance per order and spectral channel. Synthetic random noise is added via the IDL PG_RAN routine, which is appropriate for the relatively big size of our arrays ($>10^7$ elements). We then inject a planet signal containing the signal of seven molecules mixed according to the VMRs in Extended Data Table 4. This model does not represent any particular physical scenario, and abundances are chosen ad hoc to produce detectable signals from all the species included. Synthetic data are processed identically to the real data, including PCA, cross-correlation and signal estimation. It is essential to replicate the analysis sequence as closely as possible to account for any alteration of the planet signal that is due, for example, to the telluric removal procedure. Repeating the test with 30 different noise realizations always leads to a firm detection ($\sigma > 6$) of each of the seven molecular species included in the injected model. This verifies that it is possible to robustly and correctly identify molecular species in spectra containing the complex signature of seven mixed species.

As a second test, we exclude that one species can correlate with the spectrum of other species (or a combination of them). We repeat exactly the same procedure as in the previous test, except that the injected model contains the signature of six molecular species, minus the one investigated. Repeating the test with 10 noise realizations and 7 species (70 combinations) never led to a spurious detection.

As a third test, we exclude that template models containing such a dense forest of lines could just correlate, even in the absence of a signal, with the 'structure' of the Gaussian noise added to the synthetic dataset. This 'structure' arises from the fact that while the noise is randomly drawn from a Gaussian distribution, its amplitude varies, reflecting the wavelength and temporal dependence of the SNR. This causes the matrix of residual spectra after PCA to be drawn from multiple Gaussian distributions depending on the SNR of each spectral channel. We generate synthetic datasets as in the previous test, but without injecting any model, and we repeat the search for the seven molecules by cross-correlating with the same single-species models as above. We do not measure any spurious signals with significance above $4\sigma$ anywhere in the two-dimensional significance map. Furthermore, we never detect significant peaks ($>3\sigma$) at the expected planet velocity. This test was repeated ten times with consistent null results.

As a fourth test, we prove that the signal contained in the observed spectra is truly correlated in time owing to the radial velocity shift of the planet during transit. To do so, we shuffle the sequence of observed spectra in time (including the spectra out of transit) and process the shuffled dataset in the same way as the real dataset. We shuffle the spectra of each observed transit ten times for each of the six molecules detected, resulting in 60 different combinations per night. No detection at a confidence level higher than 4.5σ was detected around the nominal position of the planet (within ±30 km s$^{-1}$ from $K_P$ and ±3 km s$^{-1}$ from $V_{rest}$). Furthermore, 80% of the test yielded significance <3σ, and about 20% between 3σ and 4σ. When evaluating this outcome, it is important to realize that about 80% of our spectra are taken during transit, and so correlated signals may still be still present even after shuffling.

As a last test, despite the null detection of $CO_2$ in our data, we reject the hypothesis that any model can correlate with a time-correlated signal such as that present in our observations. This would still imply that some spectral signature from the exoplanet is present, but would of course invalidate the census of individual species. To exclude this possibility we generate two sets of synthetic models, the former containing noise randomly drawn from a Gaussian distribution (with three different variances in order to reflect the variances of our models) and the latter where the random models are skewed towards negative values in order to simulate absorption. These random models are generated at the same resolution as the physical models and then convolved with the instrumental profile of GIANO-B. The models are then interpolated to match the wavelength sampling of the instrument. As before, we consider an interval range in the CCF significance map around the nominal position of the planet rest frame (±30 km s$^{-1}$ from $K_P$ and ±3 km s$^{-1}$ from $V_{rest}$). The distribution from models containing Gaussian noise peaks around 2.2σ and has an upper boundary of 4.5σ, while the distribution from models containing skewed Gaussian noise peaks around 2.7σ and has an upper boundary of 5σ.

**Likelihood approach for model comparison**

The cross-correlation to log-likelihood mapping was performed much as described in the literature[23]. Here we summarize the main aspect of the procedure. A log-likelihood function is computed for each order, each spectrum and each radial-velocity shift of the model across the $K_P$ versus $V_{rest}$ space. To account for any alteration of the planet signal that is due to the telluric removal process, we repeat on the model the exact telluric removal process applied to the observations. The fitted telluric spectrum, stored after processing the observations, is multiplied by each transmission model tested and passed through PCA with the same number of components selected for the observations. Additionally, a high-pass filter consisting of a sliding boxcar average with a width of 120 km s$^{-1}$ is applied to both the models and the data to remove any broad-band change in the planet's effective radius as a function of wavelength. Replicating the effects of the data analysis on the models mitigates any possible bias on the likelihood analysis[23].

The variance of the data is calculated for each order and each spectrum after telluric removal and high-pass filtering. The cross-covariance between the data and the model, as well as the variance of the model, is computed for each ($K_P$, $V_{rest}$) value after post-processing the model, as explained above. Thus, our likelihood scheme implicitly assumes that the variance of each spectral channel is constant across the order. The final log-likelihood for each ($K_P$, $V_{rest}$) value is the sum of all the log-likelihoods (over each order, each night and each observed spectrum). For each model tested we obtain a likelihood map in the ($K_P$, $V_{rest}$) space. We then use a likelihood-ratio test to compare different models. For each model, we determine the maximum log-likelihood value (log$L$) around the known orbital solution of the planet (within ±6 km s$^{-1}$ from $V_{rest}$ and ±50 km s$^{-1}$ from the expected $K_P$ value). Subsequently, we apply Wilks' theorem[69] on the quantity $\Delta\log(L) = 2(\log L_1 - \log L_2)$, where log$L_1$ and log$L_2$ are the log-likelihood functions of any tested pair of models. As $\Delta\log(L)$ is distributed as a $\chi^2$ with as many degrees of freedom as parameters (5 in our case, that is metallicity, C/O, cloud fraction and the two velocities), this allows us to compute a $P$ value from $\Delta\log(L)$ and the corresponding σ-value associated with a normal distribution, as is usually done. After repeating these calculations for all the models tested, confidence intervals are defined by counting the number of standard deviations (σ) with respect to the best-fitting model. By definition this model has a $\Delta\log(L)$ of zero and therefore 0σ.

With the likelihood framework applied here, we are able to address the influence of clouds in a different but complementary way. For a model perfectly matching the data, the maximum likelihood estimator for a line-intensity scaling factor $S$ is exactly 1 (log$S$ = 0)[23]. This means that the model and data have the same average line amplitude compared to the local continuum. It is thus possible to test whether a model is a plausible representation of the data by introducing an additional parameter log$S$ and studying the value that leads to the maximum likelihood. In our case, cloud-free models are pointing to small scaling factors (−1 < log$S$ < −0.5), indicating that spectral lines are too strong compared to the observations. On the other hand, grey-cloud models computed with the current prescriptions from the literature (as described in the main text) are instead compatible at 1σ with log$S$ = 0. When the maximum-likelihood estimator $S = 1$ is adopted in our analysis, cloudy models are favoured at high significance, as shown in Fig. 3.

**Constraints on planet formation and migration processes from theoretical models**

Previous theoretical works in principle allow us to relate the atmospheric C/O ratio and metallicity of HD 209458b to its formation location and orbital evolution path, assuming that these works properly account for the physics and chemistry of protoplanetary disks as well as the main ingredients of planet formation and migration processes. Specifically, C/O ≳ 1 would be inconsistent with giant planet formation via the standard core accretion[70] beyond the H$_2$O snowline at about 2-3 au and inward migration in the protoplanetary disk with a considerable accumulation of kilometre-sized water-ice planetesimals and/or oxygen-rich gas. Indeed, this would yield sub-solar C/O ratios[12,27], that is, C/O ≲ 0.55, and super-solar metallicities. If the planet formed through core accretion outside the CO$_2$ snowline, that is, at distances greater than about 5-8 au, and then underwent disk-free migration after the disk dissipation[71], it is expected to have an atmospheric C/O ratio close to 1, in agreement with our value, and sub-solar metallicity[12].

Super-solar C/O and sub-solar metallicities are also predicted for giant planet formation via core accretion through the accumulation of millimetre- or centimetre-sized pebbles and no substantial core-envelope mixing[72], that is, the accreted solids are sequestered in the planet core. In this framework, formation beyond the H$_2$O snowline and inward migration through the protoplanetary disk would lead to C/O ≈ 0.7-0.8 and slightly sub-solar metallicity (around 0.7× to 0.8× solar), while formation outside the CO$_2$ snowline followed by disk-free migration would yield C/O ≈ 1 and lower metallicity (around 0.2× to 0.5× solar)[72]. Core erosion might release oxygen to the gaseous envelope by increasing the planet metallicity but, at the same time, decreasing the C/O ratio to solar or sub-solar values[72].

Our C/O ratio, as estimated with the assumption of thermochemical equilibrium, is therefore consistent with planet formation between the $CO_2$ and the CO snowlines and disk-free migration yielding C/O ≈ 1 and sub-solar metallicity in both the standard core accretion and pebble accretion scenarios. However, in the latter framework, pebbles are also expected to radially drift in the protoplanetary disk and transport volatile species inwards. These species would then sublimate in the proximity of snowlines, leading to a metal enrichment of the gas of the disk inside the CO snowline at about 30-40 au (refs. [11,13]). The radial drift of pebbles may thus yield a super-solar metallicity for HD 209458b as a consequence of metal-rich gas accretion. For planet formation between the $H_2O$ and $CO_2$ snowlines and subsequent migration through the protoplanetary disk, pebble drift can also increase the C/O from around 0.7-0.8 to about 0.9-1.0 (ref. [11]), and the rainout of oxygen-rich refractory species in the atmosphere can further enhance the atmospheric C/O ratio[28] when C/O < 1, which would drive the atmosphere towards the C/O ≈ 1 case we observe.

In summary, our findings would support formation of HD 209458b beyond the water snowline[11] and migration towards its host star through disk or disk-free migration. Our poor constraints on the planet metallicity do not allow us to favour theoretical scenarios predicting sub-solar or super-solar metallicities. Future observations at both low and high resolution are thus required to determine more precisely the metallicity of HD 209458b.

**Data availability**

The raw data that support the findings of this study are publicly available at the Telescopio Nazionale Galileo archive http://archives.ia2.inaf.it/tng/ hosted at the IA2 Data Center https://ia2.inaf.it/.

**Code availability**

The Gofio pipeline used to perform the GIANO-B data reduction is publicly available at https://atreides.tng.iac.es/monica.rainer/gofio. The procedures that perform the wavelength calibration, the telluric removal, the search for molecules via cross-correlation and the likelihood analysis employ public IDL libraries (explicitly indicated in the Methods) and are detailed in the text and/or in the cited papers. Even though they are available from the corresponding author upon reasonable request, we encourage other groups to develop similar tools independently and carry out their own analyses for an unbiased check of the results presented in this work. The corresponding author offers to provide any help needed. The high-resolution transmission models underpinning this article will be made available upon reasonable request to S.G. The molecular cross-sections for the various species are available on the Open Science Framework: https://osf.io/mgnw5/?view_only=5d58b814328e4600862ccfae4720acc3. The Pyrat Bay code will be released on GitHub. The radiative-equilibrium profiles computed in this article are available on Zenodo at https://zenodo.org/record/4494367.

**Acknowledgements** We thank J. Bean for comments that allowed us to improve the manuscript. P.G. gratefully acknowledges support from the Italian Space Agency (ASI) under contract 2018-24-HH.0. S.B. gratefully acknowledges support from the Italian Space Agency (ASI) under contract 2018-16-HH.0. M.B. and S.G. acknowledge support from the UK Science and Technology Facilities Council (STFC) research grant ST/S000631/1. A.S.B., G.G., A.M., G.M., and A.S. acknowledge financial contributions from the agreement ASI-INAF number 2018-16-HH.0. A.S.B., R. Claudi, G.L., A.M., V.N., L.P., A.S. and G.S. acknowledge support from PRIN INAF 2019. These results are based on observations made with the Italian Telescopio Nazionale Galileo (TNG) operated by the Fundación Galileo Galilei (FGG) of the Istituto Nazionale di Astrofisica (INAF) at the Observatorio del Roque de los Muchachos (La Palma, Canary Islands, Spain). S.N.Y. acknowledges STFC Project number ST/R000476/1. The research leading to these results received funding from the European Research Council (ERC) under the European Union's Horizon 2020 research and innovation programme (grant agreement number 679633: Exo-Atmos).

**Author contributions** P.G., M.B. and G.G. carried out the primary data reduction and data analysis. S.G. and P.C. ran theoretical models for the planet's atmosphere and transmission spectra. P.G., M.B., A.S.B., L.F., S.G., and P.C. contributed to the writing of the manuscript. P.G., M.B., S.G., A.S.B., A.S., and L.F. planned the tests to assess the reliability of the molecular detections through cross-correlation techniques. The underlying observation programme was conceived and organized by A.B., E.C., R. Claudi, S.D., S.B., A.F.L., A.M.,

E.M., G.M., I.P., E.P., G.P., and A.S. A.S.B. and V.N. planned the observations. R. Claudi is in charge of the schedules of the observations. Observations with GIANO-B were carried out by M.R., K.B., M.P., and I.C. M.R. and A.H. wrote, maintained and updated the reduction pipeline. A.B. maintained and updated the observation archive. R. Cosentino, A.F., A.H., M.P., E.P. and A.G. maintained and upgraded the GIANO-B instrument at the TNG. D.F., A.T., N.S., M.L., and A.G. contributed to the design and construction of the GIANO-B spectrograph. K.L.C. and S.N.Y. have provided molecular data. All authors contributed to the interpretation of the data and the results.

**Competing interests** The authors declare no competing interests.

**Additional information**

Correspondence and requests for materials should be addressed to P.G.

**Peer review information** Nature thanks Jacob Bean and the other, anonymous, reviewer(s) for their contribution to the peer review of this work. Peer reviewer reports are available.

Reprints and permissions information is available at [www.nature.com/reprints](www.nature.com/reprints).

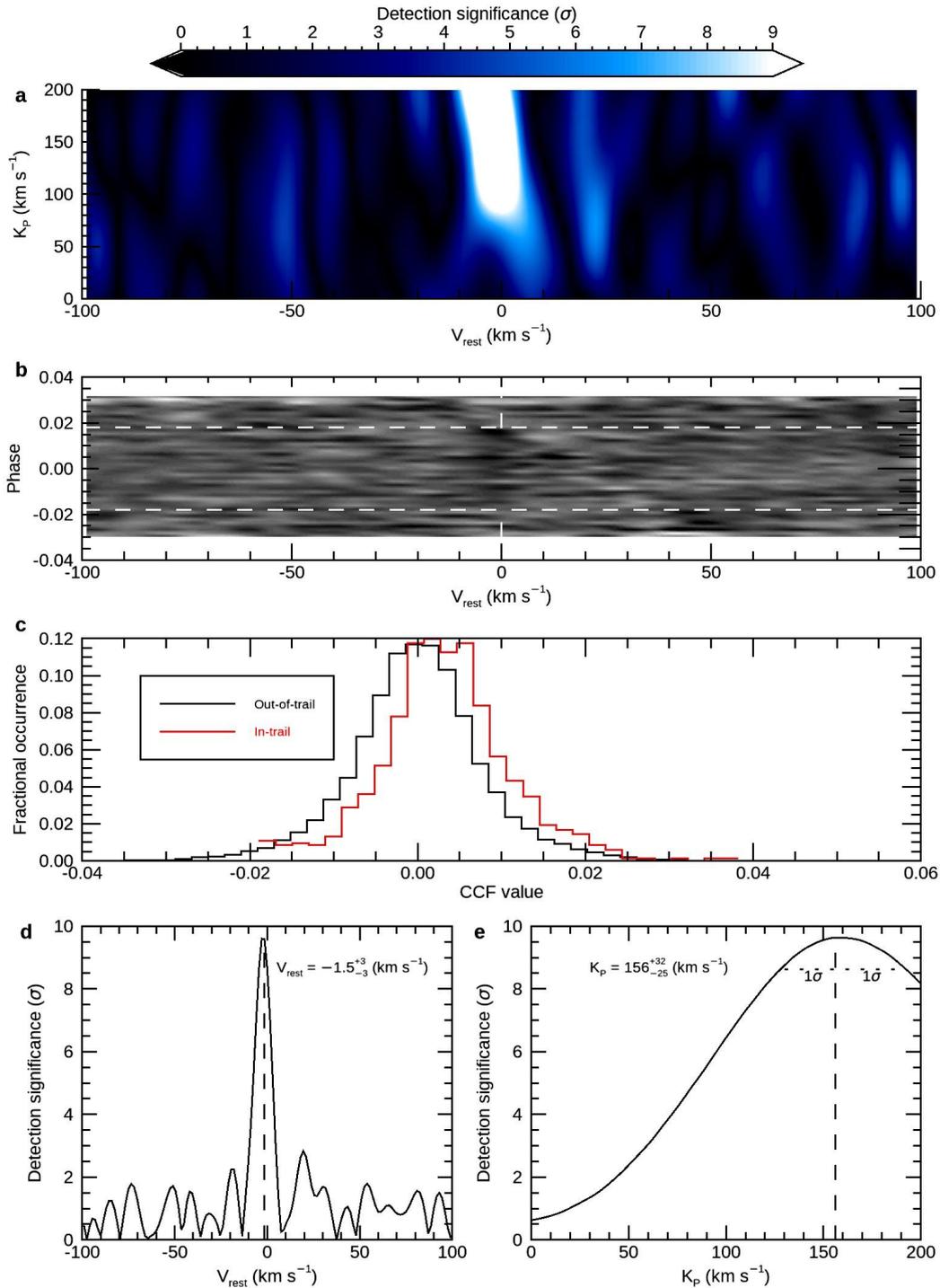

**Extended Data Fig. 1 | Detection and analysis of the H2O signal in the atmosphere of HD 209458b a**, Map of the significance of the cross-correlation values as a function of the planetary radial-velocity semi-amplitude $K_P$ and the planet rest-frame velocity $V_{rest}$. **b**, Values of the cross-correlation function in the planet rest-frame as a function of the orbital phase and $V_{rest}$. The horizontal dashed lines denote the transit ingress and egress, while the vertical dashed line indicates the expected position of the planetary signal. **c**, Distribution of CCF values in-trail and out-of-trail. **d**, Significance of the cross-correlation values as a function of $V_{rest}$. **e**, Significance of the cross-correlation values as a function of $K_P$. The dashed line indicates the peak position, while the dotted line shows the $1\sigma$ confidence interval.

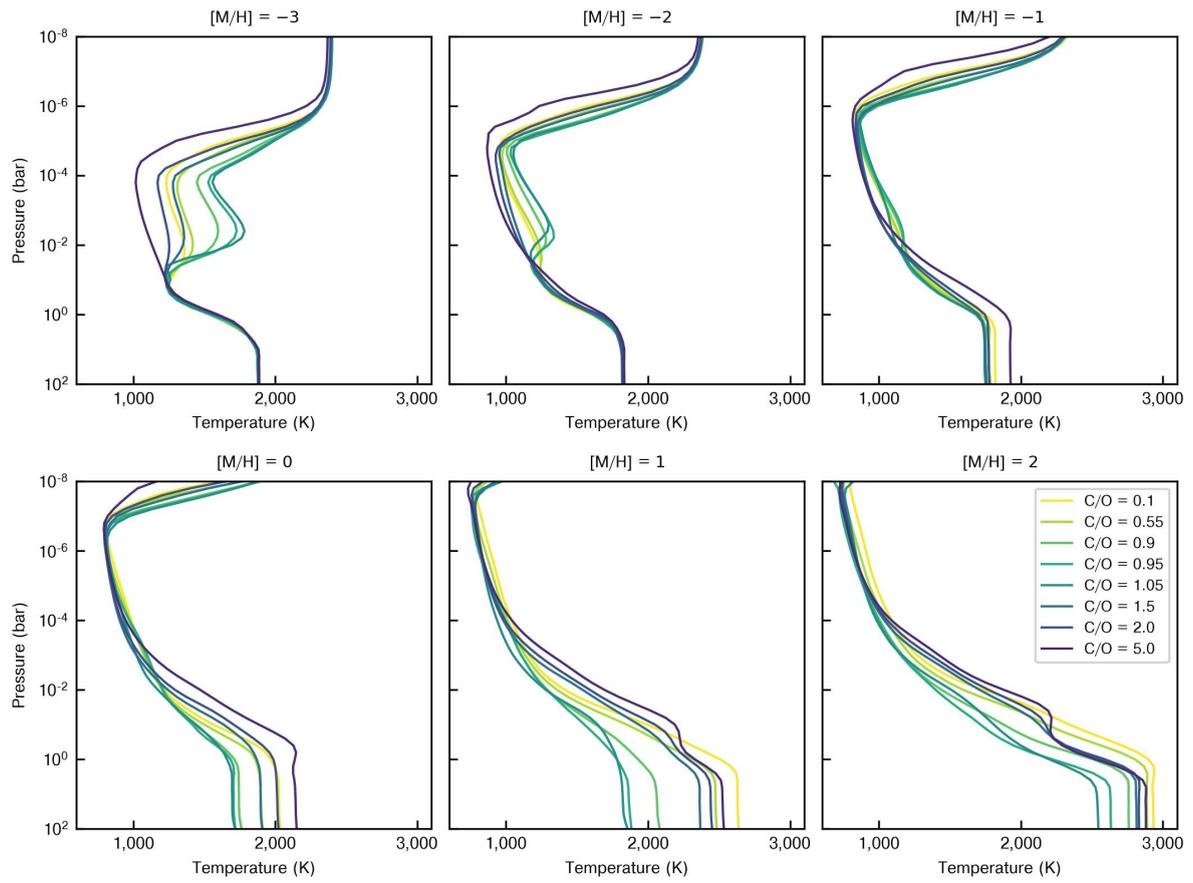

**Extended Data Fig. 2 | Theoretical pressure-temperature profiles of the atmosphere of HD 209458b** The panels show the temperature variation with pressure for different atmospheric C/O ratios and metallicities ([M/H]) under the assumption of radiative and thermochemical equilibrium.

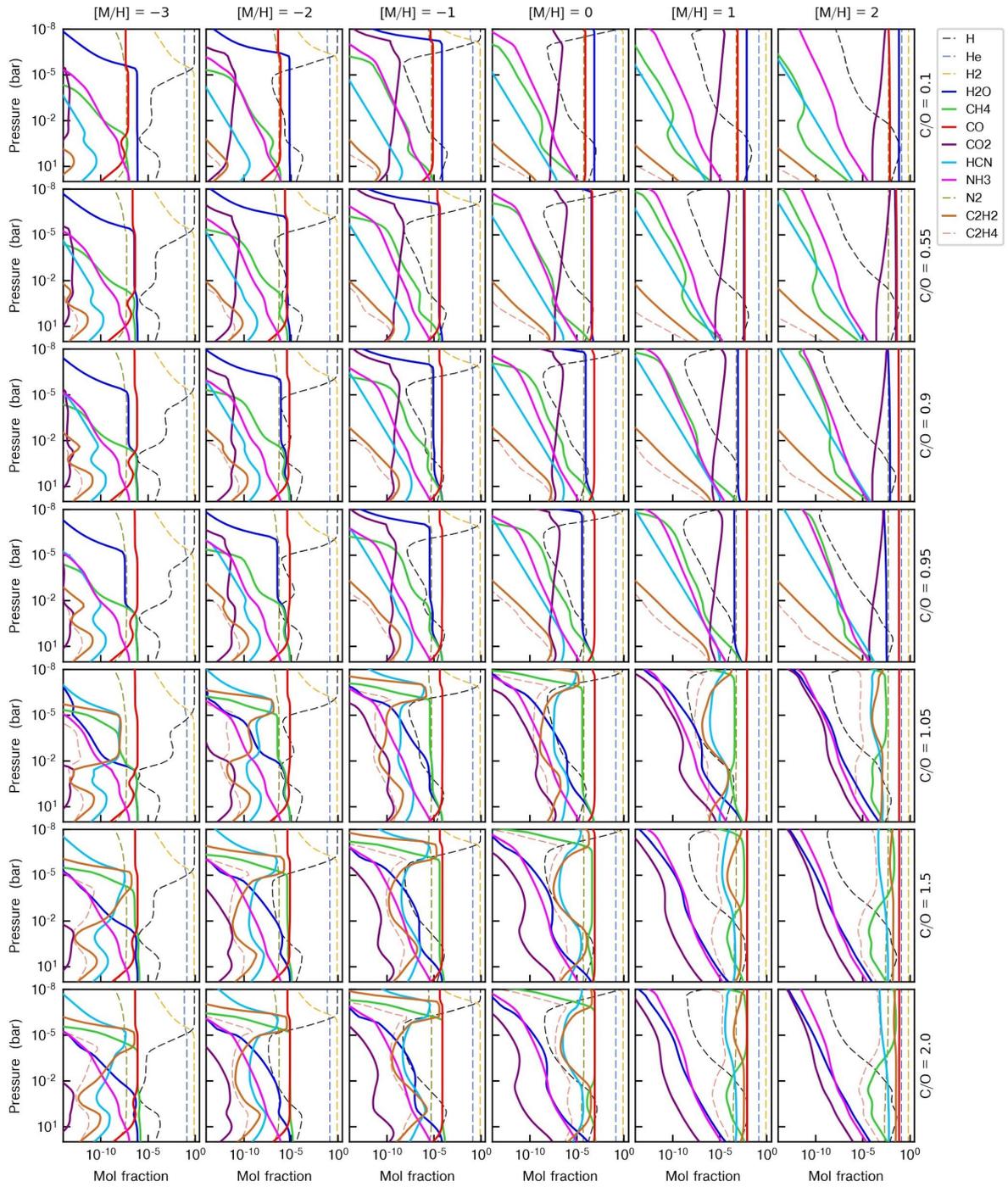

**Extended Data Fig. 3 | Theoretical pressure-abundance profiles of the atmosphere of HD 209458b.** The panels show the abundance profiles (mole fraction) of atomic and molecular species for different atmospheric C/O ratios and metallicities ([M/H]) under the assumption of radiative and thermochemical equilibrium. Each colour corresponds to a different species (see key at top-right).

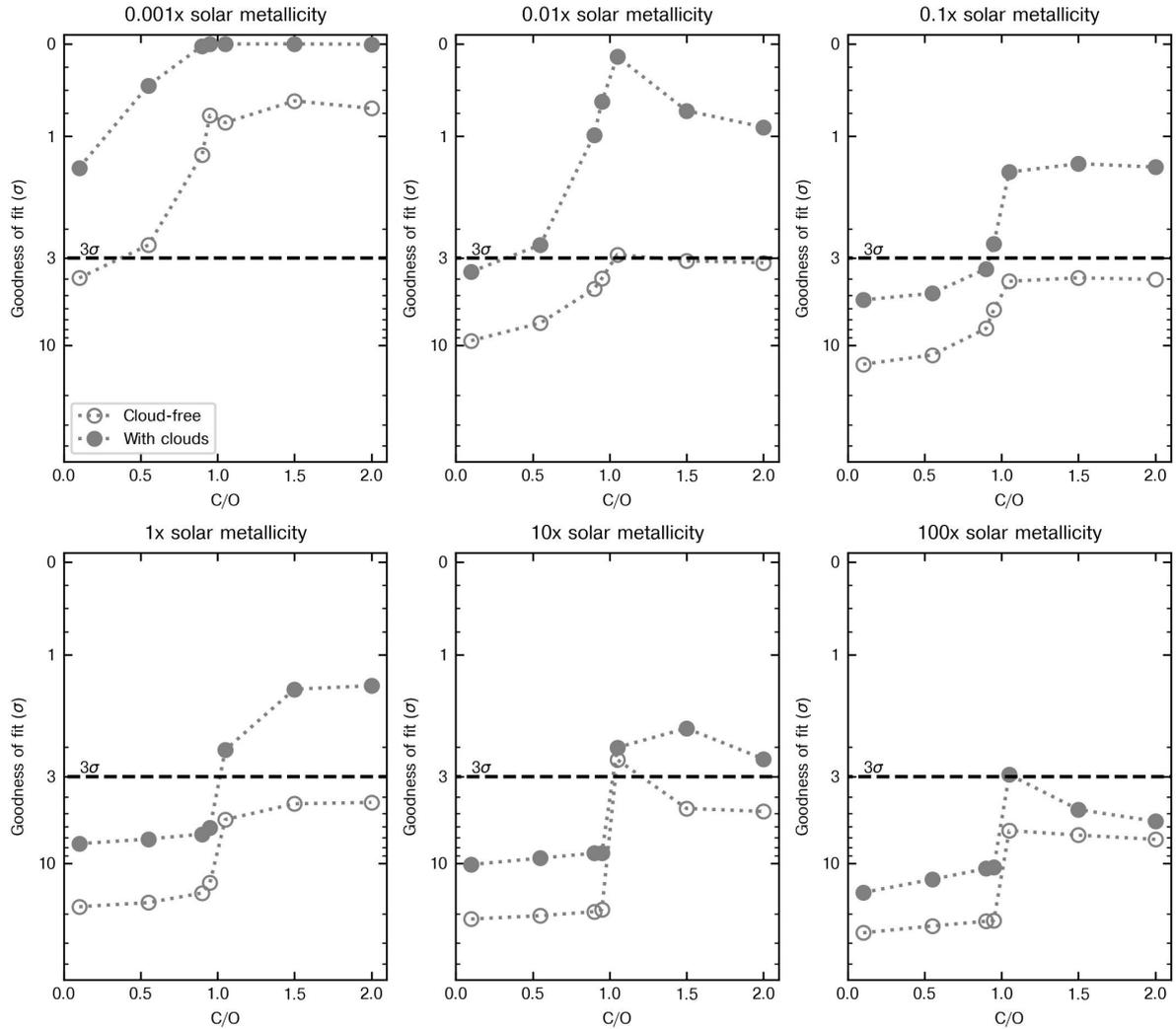

**Extended Data Fig. 4 | Comparison of atmospheric models in radiative and thermochemical equilibrium.** The six panels show the goodness of fit for the mixed models containing all the detected species as a function of C/O ratio, metallicity and presence of clouds. The filled circles represent the models with clouds, while the empty circles indicate the clear models with no clouds. The best model is found for a cloudy atmosphere with C/O = 1.05 and subsolar metallicity of 0.001 × solar (top-left panel). The goodness of fit of the models is shown with respect to the best model in units of standard deviations $\sigma$ (the higher $\sigma$, the more disfavoured the model). The horizontal dashed lines indicate the $3\sigma$ level adopted as a threshold to distinguish different scenarios. Note that for display purposes the y-axis scale is linear between $0\sigma$ and $1\sigma$, and logarithmic elsewhere.

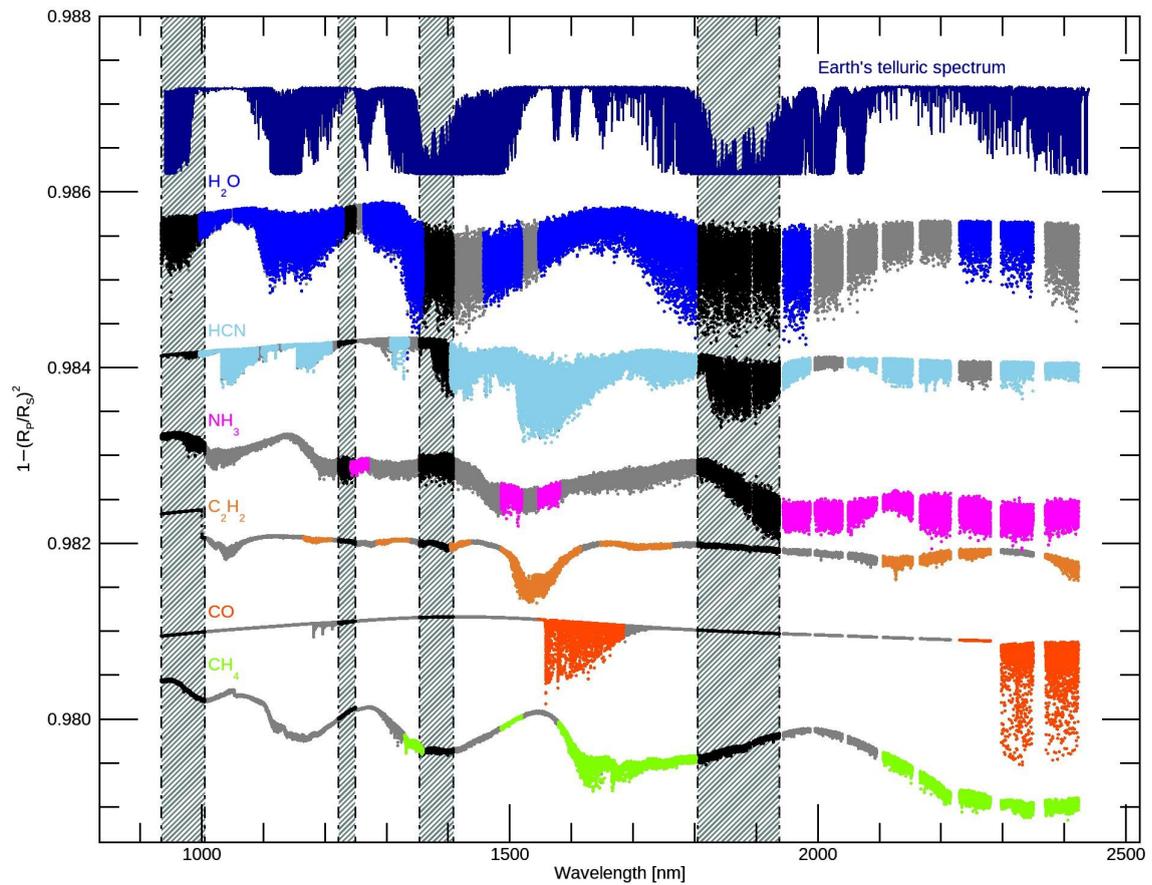

**Extended Data Fig. 5 | Selection of the GIANO-B spectral orders.** The Earth's telluric spectrum and the theoretical transmission spectra of $H_2O$, HCN, $NH_3$, $C_2H_2$, CO and $CH_4$ are shown from top to bottom in relative flux units. The colours display for each molecule the orders selected for the cross-correlation procedure, while the grey vertical bands denote the orders excluded owing to the failure of the spectral alignment and/or of the wavelength calibration procedure.

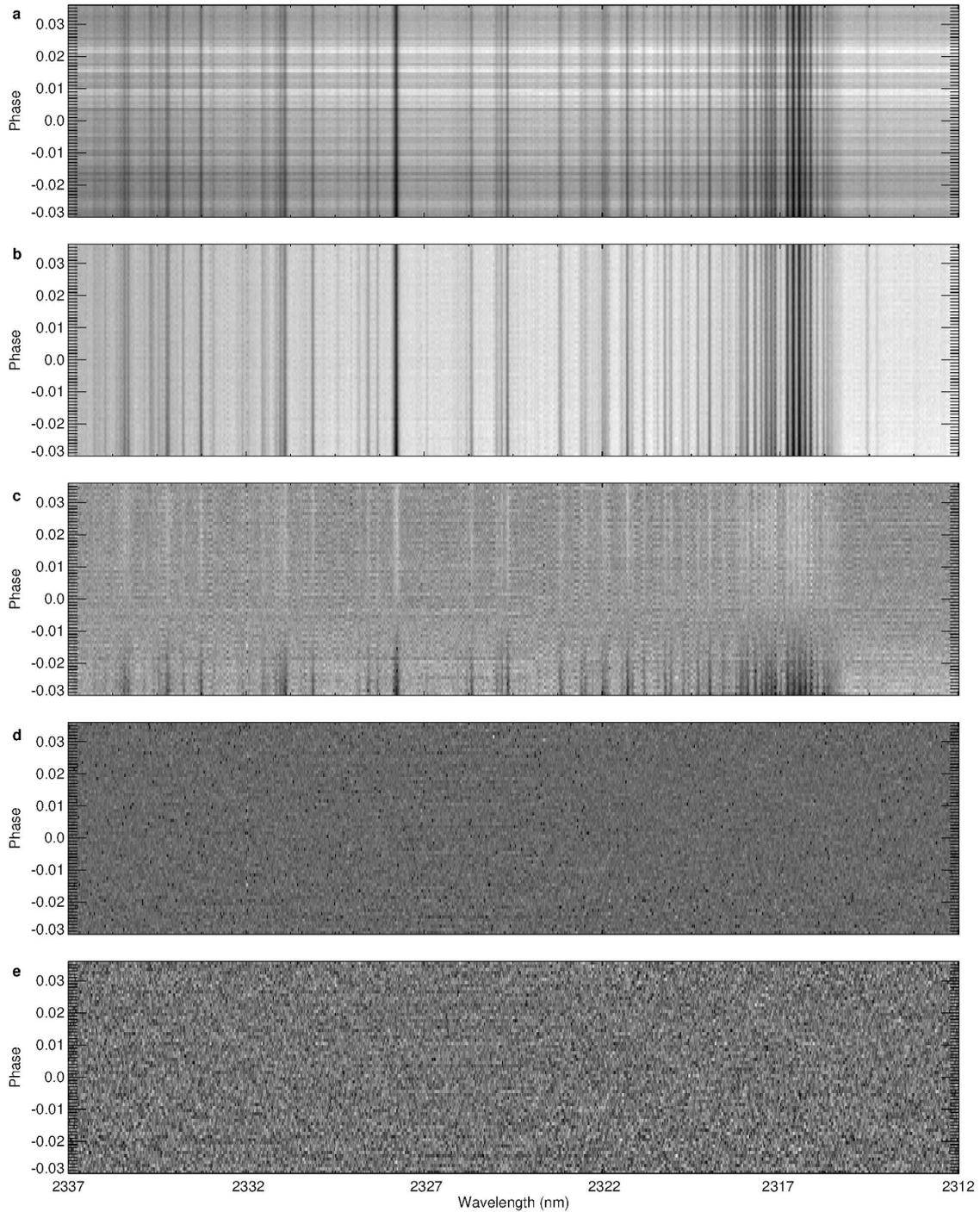

**Extended Data Fig. 6 | Stages of the analysis of GIANO-B spectra.** Example of our data reduction process over a small wavelength interval. **a**, Extracted spectra; **b**, residuals after normalization of each spectrum (each row) by its median value (throughput correction); **c**, residuals after 'standardization' of each spectral channel (each column) by mean subtraction; **d**, residuals after PCA telluric removal; **e**, residuals after division of each spectral channel by its variance and multiplication of the final spectral matrix by the median of the variances of the individual spectral channels, in order to conserve the flux (not applied in the likelihood framework).

| Night | N$_{OBS}$ | Exp. Time (s) | SNR$_{AVG}$ | SNR$_{min}$ to SNR$_{max}$ |
|---|---|---|---|---|
| 07 July 2018 | 50 | 200 | 81 | 30 - 114 |
| 29 August 2018 | 66 | 200 | 89 | 29 - 128 |
| 05 September 2018 | 26 | 200 | 112 | 7 - 178 |
| 27 August 2019 | 70 | 200 | 110 | 11 - 175 |
| 03 September 2019 | 64 | 200 | 111 | 32 - 164 |

**Extended Data Table 1 | Log of the HD 209458b GIANO-B observations**
From left to right, we report the date at the start of the night, the number of observed spectra N$_{OBS}$, the exposure time, the signal-to-noise ratio averaged across the whole spectral range SNR$_{AVG}$, and the range of signal-to-noise ratios SNR$_{min}$ to SNR$_{max}$ in the individual spectral orders.

| Parameter | Value | Reference |
|---|---|---|
| **Planetary and transit parameters** | | |
| Mid-transit time $T_c$ | 2452826.629283±0.000087 BJD | ref. 74 |
| Orbital period $P_{orb}$ | 3.52474859±0.00000038 d | ref. 74 |
| Semi-major axis $a$ | $0.04707^{+0.00045}_{-0.00047}$ au | ref. 74 |
| Transit duration $T_{14}$ | 3.072±0.003 hr | ref. 75 |
| Orbital inclination $i$ | 86.710±0.050 deg | ref. 74 |
| Orbital eccentricity $e$ | <0.0081 | ref. 68 |
| Mass $M_P$ | $0.682^{+0.014}_{-0.015}$ $M_{Jup}$ | ref. 68 |
| Radius $R_P$ | $1.359^{+0.016}_{-0.019}$ $R_{Jup}$ | ref. 68 |
| Equilibrium temperature $T_{eq}$ | 1484±18 K | ref. 76 |
| Radial-velocity semi-amplitude $K_P$ | 145 ± 1.5 km s$^{-1}$ | This work (derived from $a$, $P_{orb}$ and $i$) |
| **Stellar Parameters** | | |
| Systemic velocity $V_{sys}$ | −14.741±0.002 km s$^{-1}$ | ref. 77 |
| Effective temperature $T_{eff}$ | 6065±50 K | ref. 73 |
| Metallicity *[Fe/H]* | 0.00±0.05 dex | ref. 73 |
| Mass $M_\star$ | 1.119±0.033 $M_\odot$ | ref. 73 |
| Radius $R_\star$ | $1.155^{+0.014}_{-0.016}$ $R_\odot$ | ref. 73 |
| Age $t$ | $3.10^{+0.80}_{-0.70}$ Gyr | ref. 73 |
| Spectral type | G0V | |

**Extended Data Table 2 | HD 209458 system parameters**
Data for the planetary, transit and stellar parameters are from refs. [68,73-77].

| Species | Line list/Database | Max. Significance |
| --- | --- | --- |
| $H_2O$ | HITEMP[47] | 9.6σ |
| | POKAZATEL[43] | 8.2σ |
| $CH_4$ | HITEMP[48] | 5.6σ |
| | HITRAN[78] | 3.7σ |
| $NH_3$ | ExoMol[44] | 5.3σ |
| | HITRAN[78] | 5.1σ |
| CO | HITEMP[47,49] | 5.5σ |
| HCN | ExoMol[45] | 9.9σ |
| $C_2H_2$ | aCeTY[46] | 6.1σ |
| | ASD-1000[79] | 5.9σ |
| | HITRAN[78] | 3.3σ |
| $CO_2$ | Ames[50] | Non-detection |
| | HITEMP[47] | Non-detection |

**Extended Data Table 3 | Line list databases for the search for molecular species**
The significance in the third column refers to the cross correlation with isothermal models. Data for the line list databases are from refs. [43-50,78,79].

| Species | Line List/Database | VMR |
| --- | --- | --- |
| $H_2O$ | HITEMP[47] | $1.2 \times 10^{-4}$ |
| HCN | ExoMol[45] | $8.6 \times 10^{-5}$ |
| $CH_4$ | HITEMP[48] | $4.7 \times 10^{-3}$ |
| $NH_3$ | ExoMol[44] | $1.3 \times 10^{-4}$ |
| $C_2H_2$ | aCeTY[46] | $8.3 \times 10^{-5}$ |
| CO | HITEMP[47] | $1.4 \times 10^{-3}$ |
| $CO_2$ | HITEMP[47] | $5.3 \times 10^{-5}$ |

**Extended Data Table 4 | Line lists and VMRs of isothermal atmospheric models**
These models were used for both the cross-correlation with the GIANO-B spectra and the tests with synthetic data. Data for the line list databases are from refs. [44-48].